\documentclass[jcp,reprint,floatfix,nofootinbib,superscriptaddress]{revtex4-1}

\usepackage{amsmath, bm, amssymb}
\usepackage{float}
\usepackage{graphicx}
\usepackage{color}
\usepackage{dcolumn} 
\definecolor{cfb}{rgb}{0.0,0.9,0.9}
\definecolor{orange}{rgb}{1.0,0.45,0.0}
\definecolor{lila}{rgb}{0.7,0.0,1.0}

 \usepackage{color}
 \usepackage[normalem]{ulem}
 \usepackage{cancel}
 \newcommand{\half}{{\scriptstyle \frac{1}{2}}}
\begin{document}

\title{Spin-adapted Matrix Product States and Operators}

\author{Sebastian Keller}
\email{sebastian.keller@phys.ethz.ch}
\affiliation{ETH Z\"urich, 
Laboratory of Physical Chemistry, 
Vladimir-Prelog-Weg 2, 8093 Z\"urich, Switzerland}

\author{Markus Reiher}
\email{markus.reiher@phys.chem.ethz.ch}
\affiliation{ETH Z\"urich, 
Laboratory of Physical Chemistry, 
Vladimir-Prelog-Weg 2, 8093 Z\"urich, Switzerland}

\begin{abstract}
Matrix product states (MPSs) and matrix product operators (MPOs) allow an alternative
formulation of the density matrix renormalization group algorithm introduced by White.
Here, we describe how non-abelian spin symmetry can be exploited in MPSs and MPOs by virtue of the Wigner--Eckart theorem
at the example of the spin-adapted quantum chemical Hamiltonian operator.
\end{abstract}

\maketitle

\section{Introduction}
\label{sec:intro}

The incorporation of non-abelian symmetries
into the density matrix renormalization group (DMRG) algorithm
proposed by White\ \cite{White1992, White1993} is important
enhancing both accuracy and computational efficiency.
%
In the context of DMRG, total electronic spin symmetry, which is a non-abelian symmetry induced by the special unitary group $SU(2)$,
was first exploited in the interaction round a face (IRF) model by Sierra and Nishino \cite{Nishino1997} for quantum spin chains.
McCulloch and Gul\'acsi\ \cite{McCulloch2000, McCulloch2001, McCulloch2002} later presented a spin symmetric description in a more versatile way based on
a quasi density matrix and studied a broad range of models, including the Fermi--Hubbard model.
Their approach was subsequently adopted by Zgid and Nooijen\ \cite{Zgid2008a} and by Sharma and Chan\ \cite{Sharma2012}
to formulate a spin-adapted DMRG method for the quantum chemical Hamiltonian.
Non-abelian symmetries beyond $SU(2)$ were discussed, for instance, in Refs.\ \cite{Toth2008} and \cite{Weichselbaum2012}.
Note, however, that these more general approaches preclude the application of sum rules for Clebsch--Gordan coefficients
only available for $SU(2)$. These sum rules involve the Wigner-$6j$ and Wigner-$9j$ symbols and result in increased numerical efficiency.

Since the emergence of matrix product based DMRG\ \cite{Oestlund1995, Verstraete2004b, McCulloch2007, Crosswhite2008},
$SU(2)$ invariant matrix product states (MPSs) and matrix product operators (MPOs) for the Hamiltonian operators of the above mentioned
condensed matter models were also described\ \cite{McCulloch2007}.
These systems feature simple Hamiltonian operators with few terms compared to the quantum chemical Hamiltonian.
The spin-adaptation of the latter is a non-trivial task because all terms must first be expressed with local operators
transforming according to some irreducible representation of $SU(2)$ and subsequently be incorporated into a matrix product structure,
requiring additional coupling coefficients.

In previous work, we presented an efficient matrix product operator based formulation of the DMRG algorithm for quantum chemistry,
which we denoted second-generation DMRG\ \cite{Maquis}.
In this work, we extend our work to the development of spin-adapted MPSs and MPOs.

In Sec.\ \ref{sec:theory}, we briefly introduce the relevant formulae from quantum mechanical angular momentum
theory for the spin adaptation of MPSs and MPOs. In Secs.\ \ref{sec:mps} and \ref{sec:mpo} we demonstrate how
these formulae are applied to the construction of spin-adapted MPSs and MPOs, whereas Sec.\ \ref{sec:mps-mpo}
describes the application of MPOs to MPSs.

\section{The role of symmetries}
\label{sec:theory}

The matrix product state ansatz for a state $|\psi\rangle$ in a Hilbert space spanned by $L$ spatial orbitals
reads
\begin{equation}
| \psi \rangle = \sum_{\bm{\sigma}}
        \sum_{a_1, \ldots, a_{L-1}} \!\!\!\!\!
        M^{\sigma_1}_{1 a_1} \, M^{\sigma_2}_{a_1 a_2} \, \cdots \, M^{\sigma_L}_{a_{L-1} 1}
        \,\,
        |\bm{\sigma} \rangle,
    \label{eq:mps_long}
\end{equation}
with
 $|\bm{\sigma} \rangle =
|\sigma_1, \, \ldots, \, \sigma_L \rangle$, and
$\sigma_l = |\!\! \uparrow \! \downarrow \rangle , | \! \uparrow \, \rangle , | \! \downarrow \rangle , | \,0\, \rangle$,
which can be interpreted as a configuration interaction (CI) expansion 
where the CI coefficients are encoded as a product of matrices.

If a Hamiltonian operator possesses global symmetries, we can label its eigenstates with quantum numbers that are
the irreducible representations of the global symmetry groups.
These labels also apply to the MPS tensors $M^{\sigma_i}_{a_{i-1} a_i}$ in Eq.\ \eqref{eq:mps_long} and induce a block-diagonal structure.
For the special case of total spin symmetry, the Wigner--Eckart theorem applies, which allows us to
label the irreducible representations according to total spin (rather than to the spin projection quantum number $S_z$)
and populate the symmetry blocks with reduced matrix elements.
 
The effect of spin symmetry adaptation is therefore two-fold. Firstly, MPS tensors $M^{\sigma_i}_{a_{i-1} a_i}$ assume a block-diagonal
structure labeled by quantum numbers. Secondly, these symmetry blocks consist of reduced matrix elements obtained through
the Wigner--Eckart theorem.

\subsection{Quantum numbers}
We are interested in diagonalizing the non-relativistic electronic Coulomb Hamiltonian
\begin{equation}
  \widehat{\mathcal{H}} = \sum^L_{ij \, \sigma} t_{ij}
                    \hat{c}^\dagger_{i\sigma} \hat{c}^{\phantom{\dagger}}_{j\sigma}
                    + \frac{1}{2} \sum^L_{\substack{ijkl \\ \sigma \, \sigma'}} V_{ijkl}
                    \hat{c}^\dagger_{i\sigma} \hat{c}^\dagger_{k\sigma'}
                \hat{c}^{\phantom{\dagger}}_{l\sigma'} \hat{c}^{\phantom{\dagger}}_{j\sigma},
  \label{eq:hamil}
\end{equation}
defined on $L$ orbitals, referred to as sites.
Apart from the total spin, it conserves the particle number and point group symmetry.
We can therefore label the eigenstates with the quantum numbers $(S,N,I)$, corresponding to total spin $S$, number of electrons
$N$, and the irreducible representation $I$ of the point group of a molecule.
According to the Clebsch--Gordan expansion, we find that a composite system consisting of the two
representations $D(S_1, N_1, I_1 )$ and $ D(S_2, N_2, I_2)$ decomposes according to
\begin{align}
D(S_1, N_1, I_1 ) & \otimes D(S_2, N_2, I_2) \nonumber \\
                  & = \bigoplus_{S = |S_1-S_2|}^{S_1 + S_2} D(S, \,N_1\!+\!N_2\, , \,I_1\! \otimes\! I_2),
\label{eq:blocks}
\end{align}
where $I_1 \,\otimes \,I_2$ denotes the application of the point group multiplication table.
By employing Eq.\ \eqref{eq:blocks}, we will later be able to determine the symmetry dependent block structure of the MPS tensors.

\subsection{Reduced matrix elements}
For later reference, we introduce the required formulae for the handling of reduced matrix elements
and follow the standard treatment as presented, for example, in the book by Biedenharn and Louck \cite{biedenharn}.

Due to the fact that MPSs as well as MPOs behave like rank-$k$ tensor operators, the Wigner--Eckart theorem is
the fundamental equation to exploit spin symmetry.
It states that the matrix element of the $M$-th component $T_M^{[k]}$ of a rank-$k$ tensor operator $ \bm{T}^{[k]}$
is generated from a reduced matrix element multiplied by the Clebsch--Gordan coefficient $C^{j\,\,k\,\,j'}_{mMm'}$,
\begin{equation}
    \langle j' m' | T_M^{[k]} | jm \rangle = \langle j' || \bm{T}^{[k]} || j \rangle \,\, C^{j\,\,k\,\,j'}_{mMm'} ~.
\label{eq:wigner_eckart}
\end{equation}
The double vertical line denotes Condon and Shortley's notation for a reduced matrix element, which is independent of any projection
quantum number. We further distinguish the reduced matrix elements with bold symbols from their conventional counterparts,
a convention that we will follow throughout this work.
In the equation above, $j$ and $j'$ refer to a spin quantum number (an irreducible $SU(2)$ representation),
$m$, $m'$ and $M$ are projection quantum numbers such as the $z$-component of spin if the $z$-axis is chosen as the axis of quantization.
As the multiplet $M = -k, \ldots, k$ is determined by a single reduced matrix element, the Wigner--Eckart theorem
entails information compression allowing operators to be stored more efficiently.

For setting up the DMRG algorithm with irreducible tensor operators, it will be necessary to calculate the matrix elements of products
of tensor operators. If $S^{[k_1]}_{\mu_1}$ and $T^{[k_2]}_{\mu_2}$ are rank-$k_1$ and rank-$k_2$ tensor operators respectively,
their product will be given by
\begin{equation}
\big[S^{[k_1]} \times T^{[k_2]} \big]^{[k]}_{\mu} = \sum_{\mu_1 \mu_2} C^{j\,\,k\,\,j'}_{mMm'} S^{[k_1]}_{\mu_1} T^{[k_2]}_{\mu_2}.
\label{eq:plain_product}
\end{equation}
To benefit from information compression, we are interested in expressing the reduced matrix element of the above product
by reduced matrix elements of the individual factors.
By applying the Wigner--Eckart theorem to the product as a whole as well as to the individual elements of $\bm{S}^{[k_1]}$ and $\bm{T}^{[k_2]}$,
one obtains (for a detailed derivation see Ref.\ \cite{biedenharn})
\begin{align}
    &
    \langle j' || [\bm{S}^{[k_1]} \times \bm{T}^{[k_2]}]^{[k]} || j \rangle
    = 
    (-1)^{j+j'+k_1 + k_2}
    \nonumber\\
    &
    \quad \times
    \sum_{j''} \sqrt{(2j'' + 1) (2k + 1)}
    \left\{\begin{array}{ccc}
        j' & k_1 & j'' \nonumber \\
        k_2 & j & k
        \end{array}
    \right\}
    \nonumber\\
    &
    \qquad \times \langle j' || S^{[k_1]} || j''\rangle \,\, \langle j'' || T^{[k_2]} || j \rangle
\label{eq:6j}
\end{align}
where the quantity in curly brackets is a Wigner-$6j$ symbol.
 
If $\bm{S}^{[k_1]}$ and $\bm{T}^{[k_2]}$ act on different spaces, i.e.
\begin{align}
    \bm{S}^{[k_1]} &= \bm{S}^{[k_1]}(1) \otimes I(2), \\
    \bm{T}^{[k_2]} &= I(1) \otimes \bm{T}^{[k_2]}(2),
\end{align}
the summation over the intermediate states $j''$ in the coupling law of Eq.\ \eqref{eq:6j} can be eliminated to yield

\begin{align}
\langle j' & (j'_1 j'_2)  || [\bm{S}^{[k_1]}(1) \, \otimes \, \bm{T}^{[k2]}(2)]^{[k]} || j(j_1 j_2) \rangle =  \nonumber \\
        &\left[ \begin{array}{ccc}
                j_1 & j_2 & j \\
                k_1 & k_2 & k \\
                j'_1 & j'_2 & j'
                \end{array} 
        \right]    
        \langle j'_1 || \bm{S}^{[k_1]}(1) || j_1 \rangle \langle j'_2 || \bm{T}^{[k_2]}(2) || j_2 \rangle,
    \label{eq:9j}
\end{align}
where $j(j_1 j_2)$ means that $j_1$ and $j_2$ couple according to Eq.\ \eqref{eq:blocks} to yield $j$ and
the term in brackets is defined as the product of a Wigner-$9j$ symbol and a normalization factor,

\begin{align}
        \left[ \begin{array}{ccc}
                j_1 & j_2 & j \\
                k_1 & k_2 & k \\
                j'_1 & j'_2 & j'
                \end{array} 
        \right]
        \equiv \,\,
        & [(2j'_1 + 1)(2j'_2 + 1) (2j + 1) (2k + 1)]^{1/2} \nonumber \\
        & \times \left\{ \begin{array}{ccc}
                    j_1 & j_2 & j \\
                    k_1 & k_2 & k \\
                    j'_1 & j'_2 & j'
                    \end{array} 
                \right\}.
\label{eq:mod_9j}
\end{align}
In the subsequent sections, examples will be provided of how Eq.\ \eqref{eq:6j} and Eq.\ \eqref{eq:9j} are exploited.

\section{Symmetry-adapted MPS}
\label{sec:mps}

To understand the symmetry properties of the MPS tensors in Eq.\ \eqref{eq:mps_long},
it is important to note that the states
\begin{equation}
  |a_{l-1}\rangle = \sum_{\substack{\sigma_1, \ldots, \sigma_{l-1} \\ a_1, \ldots, a_{l-2}}} \big
                   (M^{\sigma_1}_{1 a_1} \cdots M^{\sigma_{l-1}}_{a_{l-2} a_{l-1}} \big)_{1,a_{l-1}}
                 |\sigma_1, \ldots, \sigma_{l-1} \rangle
\end{equation}  
defined on the sublattice spanned by $l-1$ sites (spatial orbitals) are mapped by $M^{\sigma_l}_{a_{l-1} a_l}$ to the states
\begin{equation}
  |a_{l}\rangle = \sum_{\sigma_l, a_{l-1}} 
                  M^{\sigma_{l}}_{a_{l-1}a_l} |a_{l-1} \rangle \otimes |\sigma_l \rangle
\end{equation}
on $l$ sites. For each value of $\sigma_l$, the MPS tensor $M^{\sigma_l}_{a_{l-1} a_l}$ therefore behaves like an \emph{operator} that maps
input states to a system enlarged by one site (spatial orbital),
where $\sigma_l$ labels the local site basis states 
$\left\lbrace |\!\! \uparrow \! \downarrow \rangle , | \! \uparrow \, \rangle , | \! \downarrow \rangle , | \,0\, \rangle \right\rbrace$,
characterized by the quantum numbers
\begin{align*}
|S, S_z, N, I\rangle = & \big\lbrace |0,0,2,A_g\rangle, \, |\frac{1}{2},\frac{1}{2},1,I\rangle, \\
                       & |\frac{1}{2},-\frac{1}{2},1,I\rangle, \, |0,0,0,A_g\rangle \big\rbrace.
\end{align*}
%
%
The operators $M^{|\uparrow \! \downarrow \rangle}$ and $M^{ |0 \rangle}$ behave like rank-$0$ tensor operators, while
$M^{| \uparrow \rangle}$ and $M^{|\downarrow \rangle}$ are the two $S_z$ components
of a rank-$\frac{1}{2}$ tensor operator.
According to the Wigner--Eckart theorem, we can therefore calculate the elements of both components from one reduced operator labeled by
total spin only,
such that the spin-adapted local basis reads
$$|S, N, I\rangle = \big\lbrace |0,2,A_g\rangle, \, |\frac{1}{2},1,I\rangle, \, |0,0,A_g\rangle. \big\rbrace $$
We now proceed as follows: from the nature of the local basis, we infer the structure of the symmetry blocks for the
complete MPS in the following section and apply the Wigner--Eckart theorem to the reduced matrix elements contained in those blocks
in Sec.\ \ref{sec:reduced_matrix_elements}.

\subsection{Symmetry blocks}
\label{sec:symmetry_blocks}

We observe that if a subsystem consisting of sites $1$ to $l-1$ is represented by states with quantum numbers $q_{l-1}$, 
the system extended to $l$ sites
will be represented by states with quantum numbers
$q_{l-1} \otimes \sigma_{l}$,
where
the tensor product for the corresponding representations is defined in Eq.\ \eqref{eq:blocks} and
$\sigma_{l}$ labels a local basis state.
If we now associate each MPS tensor $a$-index from Eq.\ \eqref{eq:mps_long} with a quantum number
\begin{equation}
    q_l = (S_l, N_l, I_l),
\label{eq:qn}
\end{equation}
each MPS tensor $\bm{M}^{\sigma_l}_{q_{l-1} a_{l-1}; q_l a_l}$ will then be characterized by the
symmetry constraint 
\begin{equation}
    q_l \in q_{l-1} \otimes \sigma_{l},
\label{eq:symmetry_constraint}
\end{equation}
which partitions the MPS tensor into symmetry blocks, indicated by the extended index $q_{l-1} a_{l-1}; q_l a_l$
supplemented with the quantum numbers $q_{l-1}$ and $q_{l}$ and separated by a semicolon for better readability.
Since they contain an $SU(2)$ irreducible representation, the corresponding MPS tensor consists of reduced matrix elements,
which we denote by a bold symbol.
Note that for an abelian symmetry, e.g., particle number, Eq.\ \eqref{eq:symmetry_constraint}
simply requires that for each block $q_{l-1}, q_l$ in $\bm{M}^{\sigma_l}_{q_{l-1} a_{l-1}; a_l q_l}$,
$N_l = N_{l-1} + N(\sigma_l)$ holds, where $N(\sigma_l)$ equals the number of particles in $\sigma_l$.
We deduce that the MPS tensor $\bm{M}^{\sigma_l}_{q_{l-1} a_{l-1}; a_l q_l}$ on site $l$ is in fact an \emph{operator}
that maps states from the subsystem spanning sites $1$ to $l-1$ to the subsystem enlarged to site $l$.
We therefore refer to $q_{l-1}$, $q_{l}$, and $\sigma_l$  as input, output, and operator quantum numbers, respectively.

The sequence of MPS tensors as they appear in Eq.\ \eqref{eq:mps_long} builds up the target state site by site
from the vacuum state.
Consequently, the quantum numbers appearing in the MPS tensors on opposite ends are the vacuum state and the target state.
By choice, we start with the vacuum state on the left hand side and finish with the target state on the right hand side of Eq.\ \eqref{eq:mps_long}.
The application of the symmetry constraint in Eq.\ \eqref{eq:symmetry_constraint} now determines which blocks will appear in the MPS tensors.
For $\bm{M}^{\sigma_1}_{q_0 a_0; q_1 a_1}$, we have one block of size $1$ denoted by
$q_0:a_0 = \lbrace (0,0,A_g):1 \rbrace$ and
\begin{equation}
q_1 : a_1 = \lbrace (0,2,A_g):1, (\frac{1}{2},1,{I_1}):1, (0,0,A_g):1 \rbrace,
\end{equation}
meaning that $\bm{M}^{\sigma_1}_{q_0 a_0; q_1 a_1}$ consists of three $1\times1$ blocks.
The MPS tensor on site $2$ shares $q_1 : a_1$ with $\bm{M}^{\sigma_1}_{q_0 a_0; q_1 a_1}$ and the output quantum numbers are
\begin{align}
q_2:a_2 = \lbrace &(0,4,A_g):1, (\half,3,I_1):1, (\half,3,I_2):1, \nonumber \\
                  & (1,2,I_1 \otimes I_2):1, (0,2,I_1 \otimes I_2):1, (0,2,A_g):2, \nonumber \\
                  & (\half,1,I_1):1, (\half,1,I_2):1, (0,0,A_g):1 \rbrace.
\end{align}
Note that the output quantum number $q_2 = (0,2,A_g)$ appears twice in the combination of the input quantum numbers with the
local site basis, namely $q_1 \otimes \sigma_2 = (0,0,A_g) \otimes (0,2,A_g)$ and $(0,2,A_g) \otimes (0,0,A_g)$.
The two blocks $q_1 \times q_2 = (0,0,A_g)\times(0,2,A_g)$ and $(0,2,A_g)\times(0,2,A_g) $ therefore have a
$1 \times 2$ shape, reflecting the fact that there are two different $(0,2,A_g)$ states defined on sites 1 and 2.
The continuation of this scheme towards the right leads to exponentially
growing block sizes, which must be limited (with the requirement that the output block sizes on site $l$ match
with the input blocks sizes on site $l+1$).
 
We further note that in this way blocks are obtained which do not appear in the set of possible blocks of the
reverse process that starts from the right hand side of the MPS by deducing the local basis states from the target quantum number.
The correct block structure is therefore obtained from the common subset of the build-up procedure from the left and the decomposition
from the right.

\subsection{Reduced matrix elements}
\label{sec:reduced_matrix_elements}

In the previous section, we established that the MPS tensor $\bm{M}^{\sigma_l}_{q_{l-1} a_{l-1}; a_l q_l}$
behaves like a set of two rank-$0$ and one rank-$\half$ irreducible tensor operator.
%
The application of the Wigner--Eckart theorem to the reduced matrix elements yields
\begin{align}
    &M^{\sigma_l}_{N_{l-1} S_{z, l-1} (k_{l-1} + a_{l-1}); \, N_{l} S_{z,l} (k_l + a_l)} \nonumber \\
    & \qquad    
        =
        \bm{M}^{\sigma_l}_{q_{l-1} a_{l-1}; q_l a_l}
        C^{S_{l-1} S_{\sigma_l} S_l}_{S_{z,l-1} m S_{z, l}},
\label{eq:mps_wigner_eckart}
\end{align}
where the blocks of the abelian MPS tensor on the left hand side of Eq.\ \eqref{eq:mps_wigner_eckart}
are labeled by pairs of the particle number $N$ and the spin projection $S_z$.
The latter may assume the values
\begin{align}
   S_{z, l-1}  &= -S_{l-1}, \ldots, S_{l-1}, \nonumber \\
   S_{z, l}    &= -S_{l}, \ldots,  S_{l}, \nonumber \\
   m           &= S_{z, l} - S_{z, l-1}. \nonumber
\end{align}
If $S_{\sigma_l}$, the spin of the local basis state $\sigma_l$, is zero, the corresponding Clebsch--Gordan coefficient
will be equal to $1$.
Note that the $a$ indices are identical on both sides of Eq.\ \eqref{eq:mps_wigner_eckart}, meaning that the reduced blocks are transferred 
as a whole and multiplied by a single Clebsch--Gordan coefficient.
In general, there is more than one reduced block on the right hand side of Eq.\ \eqref{eq:mps_wigner_eckart}
that transforms into a given block $(N_{l-1} S_{z, l-1}, N_l S_{z, l})$ on the left hand side,
such that one has to introduce pairs of row and column offsets $(k_{l-1}, k_l)$ to arrange the reduced blocks in a
block-diagonal fashion within the larger $S_z$ blocks.

Eq.\ \eqref{eq:mps_wigner_eckart} would apply if a spin-adapted MPS with reduced matrix elements
had to be transformed to the full matrix elements with abelian particle number and $S_z$ symmetry,
but not to ground state calculations, where the reduced MPS matrix elements are determined by variational optimization.

\section{Symmetry-adapted MPO}
\label{sec:mpo}

We denote the generalization of the MPS concept to MPOs as \cite{Maquis, Schollwock2011}
\begin{equation}
\widehat{\mathcal{W}} = \sum_{\bm{\sigma} \bm{\sigma'}}
    \sum_{b_1, \ldots, b_{L-1}} \!\!\!\!\!
    W^{\sigma_1 \sigma_1'}_{1 b_1}\, \cdots
    W^{\sigma_l \sigma_l'}_{b_{l-1} b_l} \cdots
    W^{\sigma_L \sigma_L'}_{b_{L-1} 1}  \,\,
    |\bm{\sigma} \rangle \langle \bm{\sigma'} |.
    \label{eq:mpo_tot}
\end{equation}
A contraction over the local site indices $\sigma_l, \sigma_l'$ in $\bm{\sigma}, \bm{\sigma'}$
leads us to define the quantities
\begin{equation}
\widehat{W}_{b_{l-1} b_l} = \sum_{\sigma_l, \sigma_l'} W^{\sigma_l \sigma_l'}_{b_{l-1} b_l}
    |\sigma_l\rangle \langle \sigma_l' |,
\label{eq:mpo_long}
\end{equation}
which are operator-valued matrices;
the entries of the
$ \widehat{W}_{b_{l-1} b_l} $ matrices are the elementary operators acting on a single site
such as the creation and annihilation
operators $\hat{c}^\dagger_{l \sigma}$ and $\hat{c}_{l \sigma}$.

\subsection{Elementary site operators}

Elementary site operators are represented by $ 4\times 4$ matrices
with respect to a basis of 
$ \lbrace |\!\! \uparrow \! \downarrow \rangle, | \! \uparrow \, \rangle, | \! \downarrow \rangle, | \,0\, \rangle \rbrace$,
e.g.,
\begin{align*}
    \hat{c}^\dagger_\uparrow = &\left( \begin{array}{cccc}
          0 & 0  & 1  & 0  \\
          0 & 0 & 0 & 1 \\
          0 & 0  & 0 & 0 \\
          0 & 0  & 0  & 0 
        \end{array} \right),
    \quad
    \hat{c}^\dagger_\downarrow = \left( \begin{array}{cccc}
          0 & -1  & 0  & 0  \\
          0 & 0 & 0 & 0 \\
          0 & 0  & 0 & 1 \\
          0 & 0  & 0  & 0 
        \end{array} \right)
    \\
    \quad\text{and} & \quad
    \widehat{F} = \left( \begin{array}{cccc}
          1 & 0  & 0  & 0  \\
          0 &-1 & 0 & 0 \\
          0 & 0  &-1 & 0 \\
          0 & 0  & 0  & 1 
        \end{array} \right), \nonumber
    \label{eq:cdagger}
\end{align*}
where $\widehat{F}$ represents the fermionic auxiliary operator to describe fermionic anticommutation (see Ref.\ \cite{Maquis}).
Note that the definition of $\hat{c}^\dagger_\downarrow$ contains a minus sign so that
${\hat{c}^\dagger_\downarrow | \uparrow \rangle = -|\!\!\uparrow\!\downarrow\rangle}$, corresponding to our choice of ordering the $\uparrow$-electron
before the $\downarrow$-electron on a single site.

To those site operators that transform according to an irreducible $SU(2)$ representation,
we may again apply the Wigner--Eckart theorem in Eq.\ \eqref{eq:wigner_eckart}.
The pairs
$\hat{c}^\dagger_\uparrow, \hat{c}^\dagger_\downarrow$ and $\hat{c}_\uparrow, \hat{c}_\downarrow$,
for instance, each form the two components of a rank-$\half$ tensor operator
with reduced matrix elements
\begin{equation}
    \hat{\bm{c}}^\dagger = \left( \begin{array}{ccc}
          0  & -\sqrt{2}   & 0  \\
          0  &  0         & 1 \\
          0  &  0         & 0 
          \end{array} \right)
\quad
    \hat{\bm{c}} = \left( \begin{array}{ccc}
          0         &  0         & 0  \\
          1         &  0         & 0 \\
          0         &  \sqrt{2}         & 0 
          \end{array} \right),
    \label{eq:cdagger}
\end{equation}
with respect to the basis $ \lbrace | 0, 2, A_g \rangle, |1, \half, I\rangle, | 0, 0, A_g \rangle \rbrace $.
The application of Eq.\ \eqref{eq:wigner_eckart} to $\hat{\bm{c}}^\dagger$ and $\hat{\bm{c}}$ yields
$\lbrace \hat{c}^\dagger_\uparrow, \hat{c}^\dagger_\downarrow \rbrace$ and $\lbrace \hat{c}_\uparrow, -\hat{c}_\downarrow \rbrace$,
respectively.

\subsection{Operator terms}

We now turn to the description of the operator terms appearing in the Hamiltonian in Eq.\ \eqref{eq:hamil}.
In analogy to the MPS case where an index with associated quantum number $q_{l-1}, a_{l-1}$ is mapped to $q_l, a_l$
by calculating the tensor product with the local site occupation $\sigma_l$, the MPO $b$ indices of Eq.\ \eqref{eq:mpo_long} may be labeled
with quantum numbers as well, where the transition from $b_{l-1}$ to $b_l$ is mediated through the action of
the local site operator located at $\widehat{W}_{b_{l-1} b_l}$.
Introducing the quantum numbers $p_{l-1}$ and $p_l$, defined according to Eq.\ \eqref{eq:qn}, we extend the notation to
$\widehat{\bm{W}}^{[k]}_{p_{l-1} b_{l-1}; p_l b_l}$,  which associates a (non-abelian) quantum number with each $b$ index
and $k$ corresponds to the rank of the elementary site operator at the location $p_{l-1} b_{l-1}; p_l b_l$.
The term
\begin{equation}
t_{ij} \hat{c}^\dagger_{i\sigma} \hat{c}^{\phantom{\dagger}}_{j \sigma}
        = t_{ij} \hat{I}_1 \otimes \ldots \otimes \hat{c}^\dagger_{i \sigma} \hat{F} \otimes \hat{F}_{i+1} \ldots
        \otimes \hat{c}_{j \sigma} \otimes \hat{I}_{j+1} \ldots
    \label{eq:2term}
\end{equation}
emerges from the repeated action of the $\widehat{\bm{W}}^{[k]}_{p_{l-1} b_{l-1}; p_l b_l}$ tensors on each site, where
the total operator quantum number is encoded in the final $p_L$ index.
We deduce that the $\widehat{\bm{W}}^{[k]}_{p_{l-1} b_{l-1}; p_l b_l}$ tensors behave like rank-$k$ tensor operators, where $k$ equals
the rank of the elementary site operator at position $b_{l-1}, b_l$.
The Wigner--Eckart theorem applies twice, first (in analogy to the MPS case) to the elements of $\widehat{\bm{W}}^{\sigma_l \sigma'_l, [k]}$.
Since the elements in this case are irreducible tensor operators themselves, the Wigner--Eckart theorem applies a second time
and yields another Clebsch--Gordan coefficient that transforms the reduced elementary site operator to full matrix elements.
In summary one obtains
\begin{align}
    &W^{\sigma_l \sigma'_l}_{N_{l-1} S_{z, l-1} I_{l-1} b_{l-1} ; N_l S_{z, l} I_l b_l} \nonumber \\ 
    & \qquad    =
        \bm{W}^{\sigma_l \sigma'_l, [k]}_{p_{l-1} b_{l-1};  p_l b_l}
        C^{S_{p_{l-1}} k S_{p_l}}_{S_{z, l-1} m S_{z,l}} C^{S_{\sigma'_l} k S_{\sigma_l}}_{S_{z, \sigma'_l} \mu S_{z, \sigma_l}}
    ,
\label{eq:redW}
\end{align}
again with the local quantum numbers
\begin{align*}
   S_{z, l-1}  &= -S_{p_{l-1}}, \ldots, S_{p_{l-1}}, \\
   S_{z, l}    &= -S_{p_l}, \ldots,  S_{p_l},        \\
   m           &= S_{z, l} - S_{z, l-1},             \\
   S_{z, \sigma'} &= -S_{\sigma'_l}, \ldots, S_{\sigma'_l}, \\
   S_{z, \sigma}  &= -S_{\sigma_l}, \ldots, S_{\sigma_l}, \\
   \mu            &= S_{\sigma_l} - S_{\sigma'_l},
\end{align*}
as before.
Note that the symmetry constraint $p_l \in p_{l-1} \otimes k$ applies and,
as the Hamiltonian operator is a spin-$0$ operator, we find that $S_{p_L} = 0$ for each term in Eq.\ \eqref{eq:hamil}.

We are now in a position to express the term in Eq.\ \eqref{eq:2term} with reduced matrix elements.
For this we need $\hat{c}^\dagger \hat{F}$ in reduced form,
\begin{equation}
    \hat{\bm{c}}^\dagger \hat{\bm{F}} = \left( \begin{array}{ccc}
          0  & \sqrt{2}   & 0  \\
          0  &  0         & 1 \\
          0  &  0         & 0 
          \end{array} \right),
\end{equation}
and the coefficients for the reduced elements of the $\widehat{\bm{W}}^{[k]}_{p_{l-1} b_{l-1}; p_l b_l}$ tensors.
The corresponding Clebsch--Gordan coefficients in Eq. \eqref{eq:redW} are all equal to $1$,
except on site $j$ where we find
\begin{equation}
    C^{\half \phantom{-}\half 0}_{\half -\half 0} = \frac{1}{\sqrt{2}}
    , \quad
    C^{\phantom{-}\half \half 0}_{-\half \half 0} = -\frac{1}{\sqrt{2}}.
\label{eq:2termclebsch}
\end{equation}
Consequently, the reduced term
\begin{equation}
    \bm{\tau}^{[\half, \half]}_{ij}
    =
    t_{ij} \,
    \sqrt{2} \,
    \hat{\bm{c}}^\dagger \! \hat{\bm{F}}_i \, \hat{\bm{c}}_j
\end{equation}
expands to the terms
$t_{ij} \hat{c}^\dagger_{i\uparrow} \hat{c}^{\phantom{\uparrow}}_{j \uparrow}$
and
$t_{ij} \hat{c}^\dagger_{i\downarrow} \hat{c}^{\phantom{\downarrow}}_{j \downarrow}$.
For the $\downarrow$-case, the minus sign from the expansion of $\hat{\bm{c}}$ to $-\hat{c}_\downarrow$
is balanced by the Clebsch--Gordan coefficient from Eq.\ \eqref{eq:2termclebsch}.
Note that since $\widehat{\bm{W}}^{[k]}_{p_{l-1} b_{l-1}; p_l b_l}$ in general contains elementary site operators
of different ranks, it does not transform irreducibly as a whole.
We may only apply the Wigner--Eckart theorem to its elements $b_{l-1}, b_l$ individually.
Further examples of reduced terms appearing in the sum of Eq.\ \eqref{eq:hamil} are given in the appendix.

\section{MPS-MPO operations}
\label{sec:mps-mpo}

\subsection{Calculations with reduced matrix elements}

We can now describe the Hamiltonian in Eq.\ \eqref{eq:hamil} and its eigenstates
with MPOs and MPSs containing reduced matrix elements
and that the latter may be transformed with Eq.\ \eqref{eq:wigner_eckart} to the MPSs and MPOs 
that we are familiar with from DMRG with abelian symmetries.
The representation based on reduced matrix elements is more efficient compared to
full matrix elements, because there are less elements to store.
In order to exploit that fact in a DMRG algorithm, however, we need to be able
to directly optimize the reduced elements in an MPS without any intermediate steps involving
the full matrix elements. 

The decisive equations of a second-generation DMRG implementation \cite{Maquis} are
the propagation of the boundaries $\mathbb{L}^{b_l}$
defined by the starting value $\mathbb{L}^{b_0 = 1}_{11} = 1$
and the recursive relation
\begin{equation}
 \mathbb{L}^{b_l}_{a_l a'_l}
            =
            \sum_{\substack{\sigma_l \sigma'_l \\ a_{l-1}, a'_{l-1} b_{l-1}}}
                N^{\sigma_l \dagger}_{a_l a_{l-1}}
                W^{\sigma_l \sigma_l'}_{b_{l-1} b_l}
                \mathbb{L}^{b_{l-1}}_{a_{l-1} a'_{l-1}}
                M^{\sigma'_l}_{a'_{l-1} a'_l}
  \label{eq:move_boundary}
\end{equation}
where the matrices $N^{\sigma_l}$ describe a second state
\begin{equation}
    | \phi \rangle = \sum_{\bm{\sigma}, a_1, \ldots, a_{L-1}}
        N^{\sigma_1}_{1 a_1} \, N^{\sigma_2}_{a_1 a_2} \, \cdots \, N^{\sigma_L}_{a_{L-1} 1}
        |\bm{\sigma} \rangle,
    \label{eq:mps2}
\end{equation}
and the matrix vector multiplication
\begin{align}
            M'^{\sigma_l}_{a_{l-1} a_l}
        =
            \sum_{\substack{\sigma'_l \\ a'_{l-1} a'_l, b_{l-1} b_l}}
                \!\!\!\!\!\!
                     W^{\sigma_l \sigma'_l}_{b_{l-1} b_l}
                     \mathbb{L}^{b_{l-1}}_{a_{l-1} a'_{l-1}}
                     M^{\sigma'_l}_{a'_{l-1} a'_l}  \mathbb{R}^{b_l}_{a'_l a_l}
    \label{eq:matrix_vector},
\end{align}
with the right boundary $\mathbb{R}^{b_l}$ defined in analogy to Eq.\ \eqref{eq:move_boundary}.
Both equations are introduced in our earlier work on second-generation DMRG \cite{Maquis}.
It is our goal to calculate the reduced matrix elements of the quantities on the left hand side
from the reduced matrix elements of the quantities on the right hand side.
Incidentally, the two equations possess the same symmetry properties, i.e., they contain the same
number of tensor operators, because $N^{\sigma_l \dagger}_{a_l a_{l-1}}$ and $\mathbb{R}^{b_l}_{a'_l a_l}$
both behave like a tensor operator whose elements obey Eq.\ \eqref{eq:wigner_eckart}.
Therefore, we only have to derive one formula for the reduced matrix elements.
In Eq.\ \eqref{eq:move_boundary}, we apply the Wigner--Eckart theorem by substituting each object
with the right hand side of Eq. \eqref{eq:wigner_eckart}, which yields
\begin{align}
    &
    \bm{\mathrm{L}}^{p_l b_l}_{q_l a_l; q'_l a'_l} C^{S_{q_l} S_{p_l} S_{q'_l}}_{m \mu m'}
    \!\!
    =
    \!\!\!\!\!\!\!\!\!
    \sum_{\substack{
                    \sigma_l \sigma'_l \\
                    a_{l-1} a'_{l-1} b_{l-1} \\
                    q_{l-1} q'_{l-1} p_{l-1}
                    }} 
    \!\!\!\!\!\!
    \sum_{\substack{m'_1 m'_2 m_1 m_2 \\ \mu_1 \mu_2}}
    \!\!\! 
    \sqrt{\frac {(2S_{q'_{l-1}}\!\!+1) (2S_{q_l}\!+1)} {(2S_{q_{l-1}}\!\!+1) (2S_{q'_l}\!+1)} }
    \nonumber\\
    &
    \times\bm{N}^{\sigma_l \dagger}_{q_l a_l; q_{l-1} a_{l-1}}
        C^{S_{q_{l-1}} S_{\sigma_l} S_{q_l}}_{m_1 m_2 m}
    \bm{W}^{\sigma_l \sigma_l', [k]}_{p_{l-1} b_{l-1}; p_l b_l}                          
        C^{S_{\sigma'_l} k S_{\sigma_l}}_{m'_2 \mu_2 m_2}                                  
        C^{S_{p_{l-1}} k S_{p_l}}_{\mu_1 \mu_2 \mu}                                  
    \nonumber\\
    &
    \times\bm{\mathrm{L}}^{p_{l-1} b_{l-1}}_{q_{l-1} a_{l-1}; q'_{l-1} a'_{l-1}}    
        C^{S_{q_{l-1}} S_{p_{l-1}} S_{q'_{l-1}}}_{m_1 \mu_1 m'_1}               
    \bm{M}^{\sigma'_l}_{q'_{l-1} a'_{l-1}; q'_l a'_l}
        C^{S_{q'_{l-1}} S_{\sigma'_l} S_{q'_l}}_{m'_1 m'_2 m'} .
\label{eq:monster}
\end{align}
Our choice of the normalization factor with the square root ensures
that Eq.\ \eqref{eq:monster} remains valid for the generation of $\bm{\mathrm{R}}^{b_l}$ 
from $\bm{\mathrm{R}}^{b_{l+1}}$.

Fortunately, we may simplify the previous equation by employing the relation \cite{biedenharn}
\begin{align}
    &
    \sum_{m'_1 m'_2 m_1 m_2 \mu_1 \mu_2}
    \Big[
    C^{S_{q_{l-1}} S_{\sigma_l} S_{q_l}}_{m_1 m_2 m}
    C^{S_{\sigma'_l} k S_{\sigma_l}}_{m'_2 \mu_2 m_2}
    \nonumber \\
    &
    \qquad
    \times
    C^{S_{p_{l-1}} k S_{p_l}}_{\mu_1 \mu_2 \mu}
    C^{S_{q_{l-1}} S_{p_{l-1}} S_{q'_{l-1}}}_{m_1 \mu_1 m'_1}
    C^{S_{q'_{l-1}} S_{\sigma'_l} S_{q'_l}}_{m'_1 m'_2 m'}
    \Big]
    \nonumber \\
    &
    \qquad \qquad
    =
    \left[
    \begin{array}{ccc}
        S_{q_{l-1}}   & S_{\sigma_l} & S_{q_l} \\
        S_{p_{l-1}} & k & S_{p_l}       \\
        S_{q'_{l-1}}  & S_{\sigma'_l} & S_{q'_l}
    \end{array} 
    \right]
    C^{S_{q_l} S_{p_l} S_{q'_l}}_{m \mu m'},
\label{eq:5c}
\end{align}
and obtain
\begin{align}
     \bm{\mathrm{L}}^{p_l b_l}_{q_l a_l; q'_l a'_l}
    &= 
        \!\!\!\!\!
        \sum_{\substack{
              \sigma_l \sigma'_l \\ 
              a_{l-1} a'_{l-1} b_{l-1} \\
              q_{l-1} q'_{l-1} p_{l-1}
             }} 
        \!\!\!\!\!
    \left[ \begin{array}{ccc}
         S_{q_{l-1}} & S_{\sigma_l} & S_{q_l} \\
         S_{p_{l-1}} & k & S_{p_l}       \\
         S_{q'_{l-1}}& S_{\sigma'_l} & S_{q'_l}
         \end{array}
    \right] 
    \nonumber\\
    &
    \times \sqrt{\frac {(2S_{q'_{l-1}}\!\!+1) (2S_{q_l}\!+1)} {(2S_{q_{l-1}}\!\!+1) (2S_{q'_l}\!+1)} }
    \quad
    \bm{N}^{\sigma_l \dagger}_{q_l a_l; q_{l-1} a_{l-1}}
    \nonumber\\
    &
    \times \bm{W}^{\sigma_l \sigma_l', [k]}_{p_{l-1} b_{l-1}; p_l b_l}                          
    \bm{\mathrm{L}}^{p_{l-1} b_{l-1}}_{q_{l-1} a_{l-1}; q'_{l-1} a'_{l-1}}    
    \bm{M}^{\sigma'_l}_{q'_{l-1} a'_{l-1}; q'_l a'_l},
\label{eq:move_boundary_reduced}
\end{align}
where the coefficient $C^{S_{q_l} S_{p_l} S_{q'_l}}_{m \mu m'}$ cancels out.
This is a remarkable result. Eq.\ \eqref{eq:move_boundary_reduced} differs from the
original version with full matrix elements [Eq.\ \eqref{eq:move_boundary}] only by the
modified Wigner-9j coupling coefficient from Eq.\ \eqref{eq:mod_9j}.
It is important to note that the Wigner-9j symbol includes the summation over all possible $S_z$ projections
of all tensor operators in Eq.\ \eqref{eq:move_boundary_reduced}.
As a consequence, the evaluation of the expectation value
$
    \langle \bm{\psi} | \sqrt{2} \, t_{ij} \, \hat{\bm{c}}^\dagger \! \hat{\bm{F}}_i \, \hat{\bm{c}}_j | \bm{\psi} \rangle
$
according to Eq.\ \eqref{eq:move_boundary_reduced} yields
\begin{equation}
      \langle \bm{\psi} | \sqrt{2} \, t_{ij} \, \hat{\bm{c}}^\dagger \! \hat{\bm{F}}_i \, \hat{\bm{c}}_j | \bm{\psi} \rangle
    =
      \langle \psi | t_{ij} \hat{c}^\dagger_{\uparrow i} \hat{c}_{\uparrow j} | \psi \rangle
    +
      \langle \psi | t_{ij} \hat{c}^\dagger_{\downarrow i} \hat{c}_{\downarrow j} | \psi \rangle.
\end{equation}
If we exchange in Eq.\ \eqref{eq:move_boundary_reduced} $\bm{N}^{\sigma_l \dagger}$ by
$\bm{\mathrm{R}}^{b_l}$, the right hand side becomes $\bm{M}'^{\sigma_l}$, the reduced
elements of the MPS tensor at site $l$ multiplied by the Hamiltonian, which is calculated
by Krylov subspace based eigensolvers such as the Jacobi--Davidson algorithm.

\subsection{The spin-adapted Hamiltonian}

In the previous section, we saw that
the two expansion products of the term in Eq.\ \eqref{eq:2termclebsch}
are both contained in the Hamiltonian in Eq.\ \eqref{eq:hamil}.
However, this is not always the case.
The group of terms
\begin{align}
    &
    \sum_{\sigma \sigma'} V_{ijji} \hat{c}^\dagger_{i \sigma} \hat{c}^\dagger_{j \sigma'} \hat{c}_{i \sigma'} \hat{c}_{\sigma j}
  =
    \nonumber \\
    & \quad
    -V_{ijji}
    \big(
        \hat{n}_{i \uparrow} \hat{n}_{j \uparrow}
      +
        \hat{n}_{i \downarrow} \hat{n}_{j \downarrow}
      +
        \hat{c}^\dagger_{i\uparrow} \hat{c}_{i\downarrow} \hat{c}^\dagger_{j\downarrow} \hat{c}_{j\uparrow}
      +
        \hat{c}^\dagger_{i\downarrow} \hat{c}_{i\uparrow} \hat{c}^\dagger_{j\uparrow} \hat{c}_{j\downarrow}
    \big)
    , 
\label{eq:flipterm}
\end{align}
%
%
%
%
for example, contains nontrivial site operators like $\hat{c}^\dagger_{i\uparrow} \hat{c}_{i\downarrow}$ and
$\hat{c}^\dagger_{j\downarrow} \hat{c}_{j\uparrow}$, which are the $S_z = 1$ and $S_z = -1$ components of a
rank-$1$ irreducible tensor operator.
In reduced form, the matrix elements of these site operators are
\begin{equation}
    \hat{\bm{c}}^\dagger \hat{\bm{c}}^{[1]} = \left( \begin{array}{ccc}
          0  &  0         & 0  \\
          0  &  \sqrt{3/2}   & 0 \\
          0  &  0         & 0 
          \end{array} \right),
\end{equation}
as the expansion to the full matrix elements [Eq.\ \eqref{eq:wigner_eckart}] confirms, because the operators
$\lbrace -\hat{c}^\dagger_\uparrow \hat{c}_\downarrow, \frac{1}{\sqrt{2}} \, (\hat{n}_\uparrow - \hat{n}_\downarrow),
        \hat{c}^\dagger_\downarrow \hat{c}_\uparrow  \rbrace $
are obtained.
If we now attempt to generate the terms in Eq.\ \eqref{eq:flipterm} from
\begin{equation}
    \bm{\tau}^{[1,1]}_{ij}
    =
    V_{ijji} \,
    \sqrt{3} \,
    \hat{\bm{c}}^\dagger \hat{\bm{c}}^{[1]}_i \hat{\bm{c}}^\dagger \hat{\bm{c}}^{[1]}_j,
    \label{eq:flipterm_su2}
\end{equation}
where the factor of $\sqrt{3}$ balances the magnitudes of the Clebsch--Gordan coefficients at site $j$ from Eq. \eqref{eq:redW},
e.g $C^{1\phantom{-}10}_{1-10}$,
we will find that there are several different possibilities of expanding the term in Eq.\ \eqref{eq:flipterm_su2} to full matrix elements.
The sum of all possibilities with a total spin of $0$ is
\begin{align}
    &
    \sum_{\substack{ \sigma_i \sigma'_i \sigma_j \sigma'_j \\ m_1 m'_1 m_2 m'_2 M_1 M_2}}
        \!\!\!\!\!
        \bm{\tau}^{[1,1]}_{ij}
        C^{ S_{\sigma'_i} 1 S_{\sigma_i} }_{m'_1 M_1 m_1}
        C^{ S_{\sigma'_j} 1 S_{\sigma_j} }_{m'_2 M_2 m_2}
        C^{110}_{M_1 M_2 0}
    =
    -V_{ijji}
    \nonumber \\
    &
    \times
    \big(
             \hat{c}^\dagger_\uparrow \hat{c}_{\downarrow i} \, \hat{c}^\dagger_\downarrow \hat{c}_{\uparrow j}
          +  \hat{c}^\dagger_\downarrow \hat{c}_{\uparrow j} \, \hat{c}^\dagger_\uparrow \hat{c}_{\downarrow i}
          +  \frac{1}{2} (\hat{n}_\uparrow - \hat{n}_\downarrow)_i \, (\hat{n}_\uparrow - \hat{n}_\downarrow)_j
    \big),
\end{align}
which does not match Eq.\ \eqref{eq:flipterm}.
For this reason, we must add a correction of $-\half \hat{\bm{n}}_i \hat{\bm{n}}_j$ to
the term in Eq.\ \eqref{eq:flipterm_su2} so that Eq.\ \eqref{eq:flipterm} is reproduced.

We have performed the analysis above for all the terms of the Hamiltonian in Eq.\ \eqref{eq:hamil},
a detailed list is provided in the appendix.

\subsection{Reduced two-site MPS tensors}

The variational optimization of two MPS sites at the same time involves
the formation of the two-site MPS tensor
\begin{equation}
P^{\sigma_l \sigma_{l+1}}_{a_{l-1} a_{l+1}} = \sum_{a_l} M^{\sigma_l}_{a_{l-1} a_l} M^{\sigma_{l+1}}_{a_l a_{l+1}}.
\end{equation}
To obtain the reduced elements of $P$, we follow the description
by Wouters {\it et al.} \cite{Wouters2014} and employ the formula in Eq.\ \eqref{eq:6j}, which couples two tensor operators and reads
\begin{align}
        &
        \bm{P}^{\sigma_l \sigma_{l+1}, [k]}_{q_{l-1} a_{l-1}; q_{l+1} a_{l+1}}
        =
        (-1)^{S_{q_{l-1}}+S_{q_{l+1}} + S_{\sigma_{l+1}} + S_{\sigma_{l}}}
        \nonumber \\
        & \quad
        \times
        \sum_{S_{q_l}} \sqrt{(S_{q_l} + 1) (2k + 1)}
        \left\{\begin{array}{ccc}
            S_{q_{l+1}} & S_{\sigma_l} & S_{q_l} \nonumber \\
            S_{\sigma_{l+1}} & S_{q_{l-1}} & k
        \end{array} \right\}
        \nonumber \\
        & \quad
        \times
        \bm{M}^{\sigma_l}_{q_{l-1} a_{l-1}; q_l a_l}  \bm{M}^{\sigma_{l+1}}_{q_l a_l; q_{l+1} a_{l+1}},
    \label{eq:6jts}
\end{align}
where $k$ runs over the expansion products of $\sigma_l \otimes \sigma_{l+1}$.
To split the two-site tensor $\bm{P}^{\sigma_l \sigma_{l+1}}$ by singular value decomposition into
$\bm{M}^{\sigma_l}$ and $\bm{M}^{\sigma_{l+1}}$, we first need to back-transform $\bm{P}^{\sigma_l \sigma_{l+1}}$
into
\begin{equation}
    \widetilde{\bm{P}}^{\sigma_l \sigma_{l+1}}_{q_{l-1} a_{l-1}; q_{l+1} a_{l+1} } = \bm{M}^{\sigma_l}_{q_{l-1} a_{l-1}; q_l a_l} \bm{M}^{\sigma_{l+1}}_{q_l a_l; q_{l+1} a_{l+1}},
\end{equation}
corresponding to the bare matrix-matrix product of $\bm{M}^{\sigma_l}$ and $\bm{M}^{\sigma_{l+1}}$.
It is given by
\begin{align}
        &
        \widetilde{\bm{P}}^{\sigma_l \sigma_{l+1}}_{q_{l-1} a_{l-1}; q_{l+1} a_{l+1}}
        =
        (-1)^{S_{q_{l-1}}+S_{q_{l+1}} + S_{\sigma_{l+1}} + S_{\sigma_{l}}}
        \nonumber\\
        & \quad \times
        \sum_{k} \sqrt{(S_{q_l} + 1) (2k + 1)}
        \left\{\begin{array}{ccc}
            S_{q_{l+1}} & S_{\sigma_l} & S_{q_l} \\
            S_{\sigma_{l+1}} & S_{q_{l-1}} & k
        \end{array} \right\}
        \nonumber \\
        & \quad \times
        \bm{P}^{\sigma_l \sigma_{l+1}, [k]}_{q_{l-1} a_{l-1}; q_{l+1} a_{l+1}}.
    \label{eq:6jts_back}
\end{align}
A singular value decomposition yields
\begin{align}
        & 
        \widetilde{\bm{P}}^{\sigma_l \sigma_{l+1}}_{q_{l-1} a_{l-1}; q_{l+1} a_{l+1}}
        \!
        =
        \!
        \sum_{a_l}
        \bm{U}^{\sigma_l}_{q_{l-1} a_{l-1}; q_l a_l}
        \bm{S}_{q_l a_l; q_l a_l}
        \bm{V}^{\sigma_{l+1}}_{q_l a_l; q_{l+1} a_{l+1}},
\label{eq:svd}
\end{align}
after which we can set
$\bm{M}^{\sigma_l} = \bm{U}, \, \bm{M}^{\sigma_{l+1}} = \bm{S} \cdot \bm{V}$
when sweeping towards the right and 
$\bm{M}^{\sigma_{l+1}} = \bm{S} \cdot \bm{V}, \, \bm{M}^{\sigma_{l+1}} = \bm{U}$
during a left sweep.
Note that, compared to Ref.\ \cite{Wouters2014}, we do not apply any normalization factor
in Eq.\ \eqref{eq:svd}.

\subsection{Reduced two-site MPO tensors}

The calculation of the reduced matrix elements of the two-site MPO tensor
\begin{equation}
\widehat{V}^{\sigma_l \sigma_{l+1} \sigma'_l \sigma'_{l+1}}_{b_{l-1} b_{l+1}}
    = \sum_{b_l} \widehat{W}^{\sigma_l \sigma_l'}_{b_{l-1} b_l} \widehat{W}^{\sigma_{l+1}\sigma'_{l+1}}_{b_l b_{l+1}}
\end{equation}
requires the application of Eq.\ \eqref{eq:6j} in analogy to the two-site MPS tensor case
to couple the matrix $b$ indices and Eq. \eqref{eq:9j} to form the tensor product of
two elementary site operators acting on separate spaces.
In summary, we obtain
\begin{align}
        & 
        \bm{V}^{\sigma_l \sigma_{l+1} \sigma'_l \sigma'_{l+1} , [k]}_{p_{l-1} b_{l-1}; p_{l+1} b_{l+1}}
        =
        (-1)^{S_{p_{l-1}}+S_{p_{l+1}} + k_1 + k_2}
        \nonumber \\
        & \quad \times
        \sum_{S_{p_l}} \sqrt{(S_{p_l} + 1) (2k + 1)}
        \left\{\begin{array}{ccc}
             S_{p_{l+1}} &      k_1    & S_{p_l} \\
                     k_2 & S_{p_{l-1}} & k
             \end{array} \right\}
        \nonumber \\
        & \quad \times
        \left[ \begin{array}{ccc}
            S_{\sigma'_l} & S_{\sigma'_{l+1}} & S_{\sigma'_l \sigma'_{l+1}} \\
            k_1           & k_2               & k                           \\
            S_{\sigma_l}  & S_{\sigma_{l+1}}  & S_{\sigma_l \sigma_{l+1}}
        \end{array} \right] 
        \bm{W}^{\sigma_l \sigma'_l, [k_1]}_{p_{l-1} b_{l-1}; p_l b_l}  \bm{W}^{\sigma_{l+1} \sigma'_{l+1}, [k_2]}_{p_l b_l; p_{l+1} b_{l+1}}.
\label{eq:6jts}
\end{align}
We emphasize, that Eq.\ \eqref{eq:move_boundary_reduced} remains valid after an exchange of $\bm{M}$ with $\bm{P}$ and $\bm{W}$ with $\bm{V}$.


\section{Numerical example}
\label{sec:example}

For illustration purposes, we compare our spin- and non-spin-adapted 
implementations in \textsc{QCMaquis} \cite{Maquis} at the example of the dioxygen molecule
and consider the lowest-lying singlet state $ ^1\Delta_g$, which is the first excited state above the triplet ground state $ ^3\Sigma_g$.
For our calculations, we employed a cc-pVTZ basis set \cite{Dunning1989} 
and correlated all 16 electrons in all 60 orbitals (full configuration interaction).
For this homonuclear diatomic, we had to adopt the highest non-abelian point group symmetry, i.e., $D_{2h}$, offered by
the {\sc Molcas} program \cite{molcas} and therefore could not consider the proper point group $D_{\infty h}$.
In our $D_{2h}$ calculations, the triplet ground state is in irreducible representation $B_{1g}$,
whereas the singlet state transforms as $A_g$ (except for the spin-contaminated calculation for $S_z$ = 0, for which we chose $C_1$). 
We compare with CCSD(T) data obtained with the same basis set from the NIST 
computational chemistry comparison and benchmark database \cite{cccbdb}. In fact, we
selected the optimized CCSD(T) internuclear distances also for the DMRG calculations (see Table\ \ref{tab:O2}).

\begin{table}[H]
\caption{O$_2$ total electronic energies for the $ ^1\Delta_g$ and $^3\Sigma_g$ states in Hartree, $E_\mathrm{H}$ (in a Dunning cc-pVTZ basis set).
'FULL' denotes that all electrons including the 1$s$ electrons were correlated. Internuclear distances are 1.22217 {\AA} and 1.20700 {\AA} for
the singlet and triplet states, respectively. 'extr.' denotes the extrapolated result.
\label{tab:O2}
}
\begin{center}
\begin{tabular}{ll}
    \hline \hline
    Method  &   $ E_{el}(^1\Delta_g$) / $E_\mathrm{H}$  \\
    \hline
    DMRG(16,60), $S_z$ = 0, $m$ = 1000      &   -150.147 369\\
    DMRG(16,60), $S$ = 0, $m$ = 1000      &     -150.114 826 \\
    DMRG(16,60), $S$ = 0, $m$ = 2000      &     -150.116 789 \\
    DMRG(16,60), $S$ = 0, $m$ = 4000      &     -150.117 657 \\
    DMRG(16,60), $S$ = 0, $m$ = 1000-4000, extr. & -150.118 164\\ 
    CCSD(T)=FULL \cite{cccbdb}   &              -150.105 829 \\
    \hline
    Method  &   $ E_{el}(^3\Sigma_g$) / $E_\mathrm{H}$  \\
    \hline
    DMRG(16,60), $S$ = 1, $m$ = 1000      & -150.151 163 \\
    DMRG(16,60), $S$ = 1, $m$ = 2000      & -150.153 533 \\
    DMRG(16,60), $S$ = 1, $m$ = 4000      & -150.154 643 \\
    DMRG(16,60), $S$ = 1, $m$ = 1000-4000, extr. & -150.155 142\\ 
    CCSD(T)=FULL \cite{cccbdb}   &  -150.153 620 \\ 
    \hline
    \hline
\end{tabular}
\end{center}
\end{table}

The non-spin-adapted energy of -150.147 369 $E_\mathrm{H}$ for the singlet state
in $C_1$ point group symmetry with $S_z = 0$ converged to the triplet state.
Hence, the singlet state will not be accessible with a non-spin-adapted algorithm,
if point group symmetry is not enforced. In the $D_{2h}$ subgroup of $D_{\infty h}$, 
the two $^1\Delta_g$ components reduce to $A_g$ and $B_{1g}$ symmetry (the $z$ axis is along the
internuclear axis) so that
considering point group symmetry would allows one to select which state is optimized.
With the spin-adapted algorithm we may set the total spin $S$ equal to zero and describe
the singlet state correctly, even if point group symmetry is not enforced.

For comparison with coupled-cluster results, we extrapolated a series of DMRG calculations with a varying number
of renormalized block states $m$ comprising
$m=1000-4000$
to calculate the singlet energy.  A conservative estimate of the accuracy is 0.5 m$E_\mathrm{H}$ given by the difference
between the extrapolated value and the best variational result with $m = 4000$.
In Table \ref{tab:O2}, we also include results from the triplet ground-state calculation for comparison.

Table\ \ref{tab:O2} clearly shows that the non-spin-adapted ($S_z=0$) result is too low in energy, whereas the correct, spin-adapted electronic 
energy of the singlet state is higher by about 0.033 $E_\mathrm{H}$. The former actually converged toward the triplet state, but the energy
for this state is not fully converged as a comparison with the converged triplet ground-state DMRG(16,60)[4000] energy in Table\ \ref{tab:O2} shows.
Note also that the CCSD(T) results are higher in energy than the corresponding DMRG results because this coupled-cluster model restricts the excitation 
operators to double
substitutions with perturbatively corrected triples. The (adiabatic) singlet-triplet gap is 125.5 kJ/mol with CCSD(T) and 97.1 kJ/mol with DMRG (for $m$=4000 as well as for the extrapolated result).
The experimental result for this gap is 94.7 kJ/mol \cite{wilk95} and therefore in very good agreement with the DMRG result. 
Note, however, that we do not include any vibrational corrections in our results.

\section{Conclusions}
\label{sec:conclusion}

Here, we developed a formalism for the incorporation of non-abelian spin symmetry into
second-generation DMRG, which is a purely MPO-based formulation of the DMRG algorithm for quantum chemistry described in Ref.\ \cite{Maquis}.
The MPO concept allows one to clearly separate the operator from the
contraction formula in which the operator is applied to a wave function.
We can therefore achieve spin adaptation for all the building blocks
consisting of elementary site-operators, the matrix product basis, and the contraction formula in Eq.\ \eqref{eq:move_boundary_reduced}
individually. This modularity facilitates a flexible implementation, which was then applied to dioxygen as a numerical example.

\section*{Acknowledgments}

This work was supported by ETH Research Grant ETH-34 12-2.

\section*{Appendix}

For reference purposes, we provide the complete list of reduced matrix elements
implemented in \textsc{QCMaquis} to represent the Hamiltonian in Eq.\ \eqref{eq:hamil}.

\subsection*{Reduced elementary site operators}
Local basis: $ \lbrace | 0, 2, A_g \rangle, |1, \half, I\rangle, | 0, 0, A_g \rangle \rbrace $.

\begin{equation}
    \hat{\bm{c}}^\dagger = \left( \begin{array}{ccc}
          0  & -\sqrt{2}   & 0  \\
          0  &  0         & 1 \\
          0  &  0         & 0 
          \end{array} \right)
\quad
    \hat{\bm{c}} = \left( \begin{array}{ccc}
          0         &  0         & 0  \\
          1         &  0         & 0 \\
          0         &  \sqrt{2}         & 0 
          \end{array} \right)
    \label{eq:cdagger}
\end{equation}

\begin{equation}
    \hat{\bm{n}} = \left( \begin{array}{ccc}
          2  &  0          & 0  \\
          0  &  1         & 0 \\
          0  &  0         & 0 
          \end{array} \right)
\quad
    \hat{\bm{c}}^\dagger \hat{\bm{c}}^{[1]} = \left( \begin{array}{ccc}
          0         &  0          & 0  \\
          0         &  \sqrt{3/2} & 0 \\
          0         &  0          & 0 
          \end{array} \right)
    \label{eq:cdagger}
\end{equation}

\begin{equation}
    \hat{\bm{c}}^\dagger\hat{\bm{n}} = \left( \begin{array}{ccc}
          0  & -\sqrt{2}   & 0  \\
          0  &  0         & 0 \\
          0  &  0         & 0 
          \end{array} \right)
\quad
    \hat{\bm{n}} \hat{\bm{c}} = \left( \begin{array}{ccc}
          0         &  0         & 0  \\
          1         &  0         & 0 \\
          0         &  0                & 0 
          \end{array} \right)
    \label{eq:cdagger}
\end{equation}

\begin{equation}
    \hat{\bm{p}}^\dagger = \left( \begin{array}{ccc}
          0  &  0          & 1  \\
          0  &  0         & 0 \\
          0  &  0         & 0 
          \end{array} \right)
\quad
    \hat{\bm{p}} = \left( \begin{array}{ccc}
          0         &  0          & 0  \\
          0         &  0          & 0 \\
          1         &  0          & 0 
          \end{array} \right)
    \label{eq:cdagger}
\end{equation}

\subsection*{Reduced Hamiltonian terms}

The definition of the two-electron integrals according to the notation in Ref.\ \cite{Knowles1989}
reads
\begin{equation}
    V_{ijkl} = \int \mathrm{d}^3r \mathrm{d}^3r' \phi^*_i({\bm{r}}) \phi^*_k({\bm{r}}') V(|{\bm{r}}-{\bm{r}}'|) \phi_j({\bm{r}}) \phi_l({\bm{r}}'),
\end{equation}
and exhibits the permutation symmetries
\begin{equation}
    V_{ijkl} = V_{klij} = V^*_{jilk} = V^*_{lkji},
\end{equation}
which give rise to equivalence classes of index permutations that share the same
two-electron integral.
We partition the Hamiltonian in Eq.\ \eqref{eq:hamil} according to these equivalence 
classes described in Table\ \ref{tab:hamterm}.
The format for the one- and two-electron integrals described in Ref. \cite{Knowles1989} only
lists unique integral values. Therefore, the second column of Table\ \ref{tab:hamterm} contains the
terms for all permutations given in the first column.
Note, that in the first column of Table\ \ref{tab:hamterm} only half of the possible two-electron index
permutations are listed in order to cancel the factor $1/2$ in Eq.\ \eqref{eq:hamil}.

\renewcommand{\arraystretch}{2}

\begin{table}
\caption{Terms of the Hamiltonian partitioned into one- and two-electron equivalence classes.}
\label{tab:hamterm}
\begin{center}
\begin{tabular}{ccc}
    \hline \hline 
    integral & terms & reduced \\
    \hline 
        $t_{ii}$
    &
        $\sum_{\sigma} \hat{c}^\dagger_{\sigma i} \hat{c}_{\sigma i}$
    &
        $\hat{\bm{n}}$
    \\
        $t_{ij} = t_{ji}$
    &
        \parbox[t]{3cm}{ \centering
            $\sum_{\sigma} \hat{c}^\dagger_{\sigma i} \hat{c}_{\sigma j}$
            \\
            $- \hat{c}_{\sigma i} \hat{c}^\dagger_{\sigma j}$
        }
    & 
        \parbox[t]{2cm}{ \centering
            $\sqrt{2} \big[ \hat{\bm{c}}^\dagger_i \hat{\bm{c}}_j \quad$
            \\
            $\quad - \hat{\bm{c}}_i \hat{\bm{c}}^\dagger_j \big]$
        }
    \\
        $V_{iiii}$
    &
        $\hat{n}_{\uparrow i} \hat{n}_{\downarrow i}$
    &
        $\hat{\bm{d}}_i$ \\
        \parbox[t]{3cm}{\centering $V_{ijjj} = V_{jijj}$}
    &
        \parbox[t]{3cm}{ \centering
            $\sum_{\sigma \neq \sigma'} \hat{c}^\dagger_{\sigma i} \hat{c}_{\sigma j} \hat{n}_{\sigma' j}$
            \\
            $- \hat{c}_{\sigma i} \hat{n}_{\sigma' i} \hat{c}^\dagger_{\sigma j}$
            \\
        }
    &
        \parbox[t]{2cm}{ \centering
            $\sqrt{2} \big[ \hat{\bm{c}}^\dagger_i \hat{\bm{n}}\hat{\bm{c}}_j \quad$
            \\
            $\quad - \hat{\bm{n}} \hat{\bm{c}}_i \hat{\bm{c}}^\dagger_j \big]$
        }
    \\
        $V_{iijj}$
    &
        $ (\hat{n}_{\uparrow i} + \hat{n}_{\downarrow i}) (\hat{n}_{\uparrow j} + \hat{n}_{\downarrow j})$
    &
        $ \hat{\bm{n}}_i \hat{\bm{n}}_j$
        \\
        \parbox[t]{3cm}{ \centering
            $V_{ijij} = V_{jiji}$
        }
    &
        \parbox[t]{3cm}{ \centering
            $ \hat{c}^\dagger_{\uparrow i} \hat{c}^\dagger_{\downarrow i} \hat{c}_{\downarrow j} \hat{c}_{\uparrow j} \quad$
            \\
            $ \quad + \hat{c}_{\downarrow i} \hat{c}_{\uparrow i} \hat{c}^\dagger_{\uparrow j} \hat{c}^\dagger_{\downarrow j}$
            \\
            $- \hat{n}_{\uparrow i} \hat{n}_{\uparrow j}$
            \\
            $- \hat{n}_{\downarrow i} \hat{n}_{\downarrow j}$
            \\
            $- \hat{c}^\dagger_{\uparrow i} \hat{c}_{\downarrow i} \hat{c}^\dagger_{\downarrow j} \hat{c}_{\uparrow j} \quad$
            \\
            $ \quad - \hat{c}^\dagger_{\downarrow i} \hat{c}_{\uparrow i} \hat{c}^\dagger_{\uparrow j} \hat{c}_{\downarrow j}$
        }
    &
        \parbox[t]{2cm}{ \centering
            $ \hat{\bm{p}}^\dagger_i \hat{\bm{p}}_j \quad$
            \\
            $ \quad + \hat{\bm{p}}_i \hat{\bm{p}}^\dagger_j$
            \\
            $- \half \hat{\bm{n}}_i \hat{\bm{n}}_j$
            \\
            $+ \sqrt{3} \,  \hat{\bm{c}}^\dagger \hat{\bm{c}}^{[1]}_i  \hat{\bm{c}}^\dagger \hat{\bm{c}}^{[1]}_j$
        }
    \\
        \parbox[t]{3cm}{ \centering
            $V_{iilk} = V_{iikl}$ =\\
            $V_{lkii} = V_{klii}$
        }
    &
        \parbox[t]{3cm}{ \centering
            $ \sum_{\sigma \sigma'} \hat{n}_{\sigma i} \hat{c}^\dagger_{\sigma' k} \hat{c}_{\sigma' l} $
            \\
            $- \sum_{\sigma \sigma'} \hat{n}_{\sigma i} \hat{c}_{\sigma' k} \hat{c}^\dagger_{\sigma' l} $
            \\
        }
    &
        \parbox[t]{2cm}{ \centering
            $\sqrt{2} \hat{\bm{c}}^\dagger \hat{\bm{n}}_i \hat{\bm{c}}^\dagger_k \hat{\bm{c}}_l$
            \\
            $-\sqrt{2} \hat{\bm{c}}^\dagger \hat{\bm{n}}_i \hat{\bm{c}}_k \hat{\bm{c}}^\dagger_l$
        }
    \\
        \parbox[t]{3cm}{ \centering
            $V_{ijil} = V_{ijli}$ =\\
            $V_{jiil} = V_{liji}$
        }
    &
        \parbox[t]{3cm}{ \centering
            $ \sum_{\sigma \neq \sigma'} \hat{c}^\dagger_{\sigma} \hat{c}^\dagger_{\sigma' i} \hat{c}_{\sigma' j} \hat{c}_{\sigma l}$
            \\
            $ \sum_{\sigma \neq \sigma'} \hat{c}_{\sigma} \hat{c}_{\sigma' i} \hat{c}^\dagger_{\sigma' j} \hat{c}^\dagger_{\sigma l}$
            \\
            $- \sum_{\sigma \sigma'} \hat{c}^\dagger_{\sigma} \hat{c}_{\sigma' i} \hat{c}^\dagger_{\sigma' j} \hat{c}_{\sigma l}$
            \\
            $+ \sum_{\sigma \sigma'} \hat{c}^\dagger_{\sigma} \hat{c}_{\sigma' i} \hat{c}_{\sigma' j} \hat{c}^\dagger_{\sigma l}$
        }
    &
        \parbox[t]{2cm}{ \centering
            $ -\sqrt{2} \hat{\bm{p}}^\dagger_i \hat{\bm{c}}_j \hat{\bm{c}}_l$
            \\
            $ -\sqrt{2} \hat{\bm{p}}_i \hat{\bm{c}}^\dagger_j \hat{\bm{c}}^\dagger_l$
            \\
            $ \sqrt{3} \hat{\bm{c}}^\dagger \hat{\bm{c}}^{[1]}_i \hat{\bm{c}}^\dagger_j \hat{\bm{c}}_l$
            \\
            $ -\half \sqrt{2}\hat{\bm{n}}_i \hat{\bm{c}}^\dagger_j \hat{\bm{c}}_l$
            \\
            $ +\sqrt{3} \hat{\bm{c}}^\dagger \hat{\bm{c}}^{[1]}_i \hat{\bm{c}}_j \hat{\bm{c}}^\dagger_l$
            \\
            $ +\half \sqrt{2}\hat{\bm{n}}_i \hat{\bm{c}}_j \hat{\bm{c}}^\dagger_l$
        }
    \\
        \parbox[t]{3cm}{ \centering
            $V_{ijkl} = V_{ijlk}$ =\\
            $V_{jikl} = V_{jilk}$
        }
    &
        \parbox[t]{3cm}{ \centering
            $  \sum_{\sigma \sigma'} \hat{c}^\dagger_{\sigma i} \hat{c}^\dagger_{\sigma' k} \hat{c}_{\sigma' l} \hat{c}_{\sigma j}$
            \\
            $+ \sum_{\sigma \sigma'} \hat{c}^\dagger_{\sigma i} \hat{c}^\dagger_{\sigma' l} \hat{c}_{\sigma' k} \hat{c}_{\sigma j}$
            \\
            $+ \sum_{\sigma \sigma'} \hat{c}^\dagger_{\sigma j} \hat{c}^\dagger_{\sigma' k} \hat{c}_{\sigma' l} \hat{c}_{\sigma i}$
            \\
            $+ \sum_{\sigma \sigma'} \hat{c}^\dagger_{\sigma j} \hat{c}^\dagger_{\sigma' l} \hat{c}_{\sigma' k} \hat{c}_{\sigma i}$
        }
    &
        \parbox[t]{2cm}{ \centering
            \emph{see Table\ \ref{tab:4term}}
        }
    \\
    & & \\
    \hline
\end{tabular}
\end{center}
\end{table}

\begin{table}[H]
\caption{\emph{continued from Table \ref{tab:hamterm}, lower right}. Reduced Hamiltonian terms for the case $i \neq j \neq k \neq l$.
}
\label{tab:4term}
\begin{center}
\begin{tabular}{cc}
    \hline \hline 
    & reduced \\
    \hline
    &
        \parbox[t]{7cm}{ \centering
            $ \phantom{+} \text{sgn}(\pi_{ijkl})
              \big[
                 \alpha \big(\hat{\bm{c}}_{j} \hat{\bm{c}}_{l} \big)^{[1]} \hat{\bm{c}}^\dagger_{k} \hat{\bm{c}}^\dagger_{i}
                + \beta \big(\hat{\bm{c}}_{j} \hat{\bm{c}}_{l} \big)^{[0]} \hat{\bm{c}}^\dagger_{k} \hat{\bm{c}}^\dagger_{i}
              \big]$
            \\
            $ + \text{sgn}(\pi_{ijlk})
              \big[
                 \alpha \big(\hat{\bm{c}}_{j} \hat{\bm{c}}^\dagger_{k} \big)^{[1]} \hat{\bm{c}}_{l} \hat{\bm{c}}^\dagger_{i}
                + \beta \big(\hat{\bm{c}}_{j} \hat{\bm{c}}^\dagger_{k} \big)^{[0]} \hat{\bm{c}}_{l} \hat{\bm{c}}^\dagger_{i}
              \big]$
            $ + \text{sgn}(\pi_{jikl})
              \big[
                 \alpha \big(\hat{\bm{c}}^\dagger_{i} \hat{\bm{c}}_{l} \big)^{[1]} \hat{\bm{c}}^\dagger_{k} \hat{\bm{c}}_{j}
                + \beta \big(\hat{\bm{c}}^\dagger_{i} \hat{\bm{c}}_{l} \big)^{[0]} \hat{\bm{c}}^\dagger_{k} \hat{\bm{c}}_{j}
              \big]$
            \\
            $ + \text{sgn}(\pi_{jilk})
              \big[
                 \alpha \big(\hat{\bm{c}}^\dagger_{i} \hat{\bm{c}}^\dagger_{k} \big)^{[1]} \hat{\bm{c}}_{l} \hat{\bm{c}}_{j}
                + \beta \big(\hat{\bm{c}}^\dagger_{i} \hat{\bm{c}}^\dagger_{k} \big)^{[0]} \hat{\bm{c}}_{l} \hat{\bm{c}}_{j}
              \big]$
            \\
        }
    \\        
        \parbox[t]{2cm} {\centering
            $k > l$, $l > j$
        }
    &
        $\alpha = -\sqrt{3}$, $\beta = 1$
    \\
        \parbox[t]{2cm} {\centering
            $k > j$, $j > l$
        }
    &
        $\alpha = -\sqrt{3}$, $\beta = -1$
    \\
        \parbox[t]{2cm} {\centering
            $j > k$, $k > l$
        }
    &
        $\alpha = 0$, $\beta = 2$
    \\
     & \\
    \hline
\end{tabular}
\end{center}
\end{table}

\end{document}